\documentclass{article}
\usepackage{ijcai25}

\usepackage{times}
\usepackage{soul}
\usepackage{url}
\usepackage[hidelinks]{hyperref}
\usepackage[utf8]{inputenc}
\usepackage[small]{caption}
\usepackage{graphicx}
\usepackage{amsmath}
\usepackage{amsthm}
\usepackage{amsthm}
\usepackage{amssymb}
\usepackage{booktabs}
\usepackage{algorithm}
\usepackage{algorithmic}
\usepackage[switch]{lineno}
\usepackage{subfigure}
\usepackage{comment}

\title{Learning to Coordinate Under Threshold Rewards: A Cooperative Multi-Agent Bandit Framework}

\author{
Michael Ledford$^1$
\and
William Regli$^1$\\
\affiliations
$^1$University of Maryland, College Park\\
\emails
\{mledfor, regli\}@umd.edu
}

\begin{document}

\maketitle

\begin{abstract}
    Cooperative multi-agent systems often face tasks that require coordinated actions under uncertainty. While multi-armed bandit (MAB) problems provide a powerful framework for decentralized learning, most prior work assumes individually attainable rewards. We address the challenging setting where rewards are \textit{threshold-activated}: an arm yields a payoff only when a minimum number of agents pull it simultaneously, with this threshold unknown in advance. Complicating matters further, some arms are decoys—requiring coordination to activate but yielding no reward—introducing a new challenge of wasted joint exploration. We introduce Threshold-Coop-UCB (T-Coop-UCB), a decentralized algorithm that enables agents to jointly learn activation thresholds and reward distributions, forming effective coalitions without centralized control. Empirical results show that T-Coop-UCB consistently outperforms baseline methods in cumulative reward, regret, and coordination metrics, achieving near-Oracle performance. Our findings underscore the importance of joint threshold learning and decoy avoidance for scalable, decentralized cooperation in complex multi-agent environments.

\end{abstract}

\section{Introduction}

Cooperative multi-agent systems are increasingly central to real-world decision-making in domains such as disaster response, collaborative robotics, and distributed sensing---scenarios where agents must coordinate under uncertainty to accomplish tasks that cannot be completed independently. While some environments support flexible or ad hoc interaction~\cite{stone2010ad,barrett2014communicating}, others require structured coordination, such as UAV swarms jointly lifting a payload or maintaining persistent sensor coverage.

The uncertainty, partial observability, and decentralized nature of these environments make multi-agent multi-armed bandit (MA-MAB) problems a natural framework for studying distributed learning and coordination. Unlike classical MABs, which focus solely on balancing exploration and exploitation, the multi-agent setting adds challenges: agents must coordinate actions, share information, and reason about joint outcomes.

Although classical bandit algorithms have been extensively studied~\cite{gittins1979bandit,auer1995gambling,lai1985asymptotically,auer2002finite}, naively applying them individually across agents in cooperative settings often leads to suboptimal outcomes. Prior work has explored decentralized exploration~\cite{lalitha2021bayesian}, reward sharing~\cite{landgren2018social}, and coordination under communication constraints~\cite{chakraborty2017coordinated,chang2023optimal}, but most approaches assume that rewards are independently triggered or that coordination requirements (e.g., task structure, coalition size) are known in advance.

In contrast, we study a setting where coordination is \textit{ambiguous and must be learned through interaction}. We introduce the problem of \textit{threshold-activated rewards}, where an arm yields a payoff only if a coalition of agents simultaneously selects it and meets an unknown activation threshold. This setting captures a broad class of real-world coordination problems, where success depends not only on selecting the right task, but doing so with the right number of collaborators. It violates standard assumptions in prior MA-MAB work, which rely on observable individual feedback or known coordination structure.

To further compound exploration, we introduce \textit{decoy arms}---actions that require joint activation but yield no reward. These arms deliberately mislead agents into expending coordinated effort on unproductive options, introducing a new and realistic failure mode in multi-agent settings. Unlike prior work on capacity-limited or collision-avoidance settings~\cite{wang2022multi,liu2010distributed}, our environment requires intentional convergence on actions for any reward to occur, with the added challenge that \textit{some coordinated efforts are intentionally deceptive}.

Despite their relevance to domains like multi-robot teaming, distributed task allocation, and human-agent collaboration, \textit{threshold-activated rewards and decoy arms remain underexplored} in the cooperative bandit literature. These challenges raise a fundamental question: \textit{How can decentralized agents learn not only which actions are valuable, but how many collaborators are needed---and which tasks are worth the effort to begin with?}

While our approach primarily addresses game-theoretic coordination under uncertainty, it also shares conceptual links with social choice theory, particularly regarding how groups can efficiently coordinate on mutually beneficial decisions without centralized guidance. Such generalized MAB models have direct applications in collaborative robotics, disaster response, and distributed sensor networks, where uncertain thresholds and joint coordination frequently arise.

To address this, we propose \textbf{Threshold-Coop-UCB (T-Coop-UCB)}, a decentralized algorithm that enables agents to jointly estimate both activation thresholds and reward distributions while forming dynamic coalitions to maximize cumulative reward. T-Coop-UCB handles ambiguous zero-reward feedback, requires no centralized control or prior task knowledge, and balances distributed exploration with threshold-sensitive coordination.

\textbf{Our key contributions are:}
\begin{itemize}
    \item We introduce a novel MA-MAB framework with unknown threshold-activated rewards and adversarial decoy arms that mislead coordination.
    \item We propose \textbf{T-Coop-UCB}, a decentralized algorithm that jointly learns both coalition thresholds and reward estimates while coordinating under ambiguity.
    \item We show empirically that T-Coop-UCB consistently outperforms strong baselines across cumulative reward, regret, and coordination metrics---achieving near-Oracle performance.
\end{itemize}

\begin{figure}[t]
    \centering
    \includegraphics[width=0.8\linewidth]{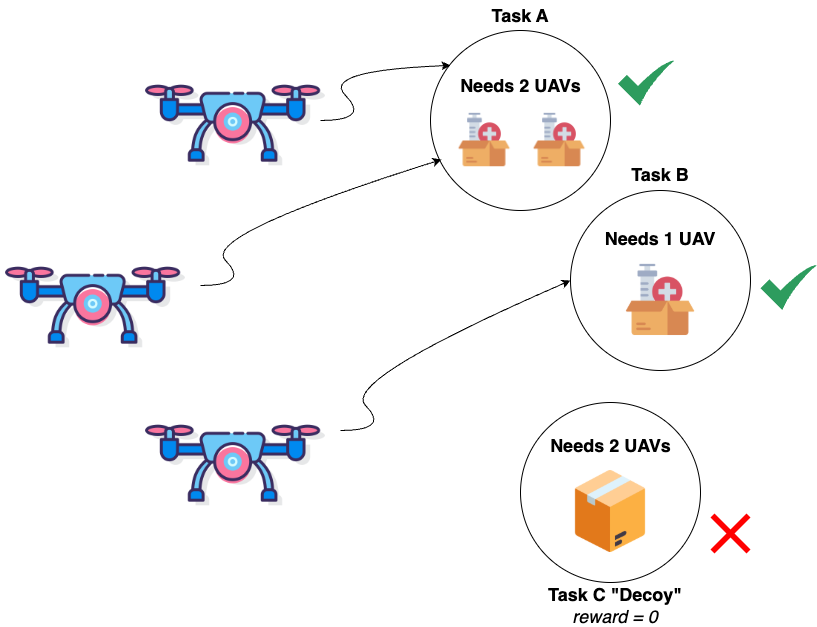} 
    \caption{\textbf{UAV Coordinated Task.} A team of three UAVs must choose among tasks with varying coordination requirements. Some tasks require multiple UAVs to activate a reward, while one is a decoy offering no payoff. Agents must learn which tasks are valuable and how many teammates are needed for success.}

    \label{fig:uav_coordination}
\end{figure}

Figure~\ref{fig:uav_coordination} illustrates a representative scenario where a team of UAVs must discover effective coordination patterns, identify decoys, and adaptively allocate themselves to tasks with unknown cooperation requirements.


\section{Related Work}

Multi-armed bandit (MAB) problems have been extensively studied as models for online decision-making under uncertainty. Classical work addresses the exploration-exploitation trade-off in single-agent settings~\cite{gittins1979bandit,auer2002finite,lai1985asymptotically}, introducing foundational algorithms such as \textit{Upper Confidence Bound (UCB)} and \textit{Thompson Sampling}. These approaches typically assume full observability and centralized, single-agent learning.

\paragraph{Decentralized and Cooperative Multi-Agent Bandits.}
Multi-agent extensions of MABs model scenarios where agents operate concurrently in a shared environment, either learning independently~\cite{lattimore2020bandit} or cooperatively through information sharing~\cite{landgren2018social}. Prior work has addressed decentralized exploration~\cite{lalitha2021bayesian}, distributed coordination~\cite{kalathil2014decentralized}, and reward-sharing under communication constraints~\cite{landgren2021distributed}. However, unlike our setting, most models assume that rewards are individually triggered, and do not require coordinated joint actions to receive feedback.

\paragraph{Collision-Avoidance in Decentralized Bandits.}
Another line of work focuses on avoiding simultaneous actions that lead to interference or \textit{collisions}, particularly in cognitive radio and distributed resource allocation settings \cite{liu2010distributed}. Algorithms such as \textsc{MusicalChairs}~\cite{rosenski2016multi}, \textsc{SIC-MMAB}~\cite{boursier2019sic}, and \textsc{GameofThrones}~\cite{bistritz2021game} are designed to minimize overlap among agents by assigning distinct arms, rather than promoting collaboration. In contrast, our setting requires agents to intentionally coordinate their actions to trigger rewards. This makes collision a necessary feature rather than a failure mode.

\paragraph{Collaborative and Communication-Aware Learning.}
Some work explores communication-aware bandit settings, where agents reason about when and how to communicate under bandwidth or cost constraints~\cite{chakraborty2017coordinated,agarwal2022multi,martinez2018decentralized,chang2023optimal}. These models focus on optimizing the trade-off between learning speed and communication cost. Our work assumes free communication but emphasizes the more fundamental challenge of learning \textit{how to coordinate}, rather than \textit{when to communicate}. We introduce a setting where coordination itself is ambiguous and must be discovered through experience.

\paragraph{Threshold-Activated and Structured Bandits.}
Our work is most closely related to bandit models with threshold-based feedback. Abernethy et al.~\shortcite{abernethy2016threshold} study a single-agent threshold bandit problem where a reward is received only if an arm's \textit{expected value} exceeds a fixed threshold. Their setting models a threshold value as contextual side information that the agent knows in advance and uses it for decision-making. In contrast, we consider a setting with multi-agent threshold-activated rewards, where the \textit{threshold} refers to a minimum number of agents required to jointly activate the arm---and this threshold is unknown and must be learned. Unlike Abernethy et al.~\shortcite{abernethy2016threshold}, our agents must learn both the threshold and the reward distribution through joint exploration.

Magesh et al.~\shortcite{magesh2021decentralized} study decentralized learning with non-zero collisions, where agents receive partial rewards even when colliding on the same arm, up to a known threshold $N$. In their setting, exceeding the threshold cancels the reward, but the agents know the value of $N$ in advanced. Our setting is fundamentally different: rewards are probabilistic and only triggered if a hidden activation threshold is satisfied. Moreover, sending more agents than necessary yields no extra benefit but does not cancel the reward---making our setting more aligned with real-world collaborative tasks like transport or surveillance, where redundancy does not prevent success but wastes opportunity.

\paragraph{Exploration under Unknown Coordination Requirements.}
Wang et al.~\shortcite{wang2022multi} study a related problem in wireless networks and edge computing, where agents interact with arms that have finite shareable resources. In their model, the total reward scales with the number of agents pulling an arm, up to a known or learnable capacity. A single agent pulling an arm still receives a proportional reward, and the primary challenge is to avoid oversaturating arms while maximizing total reward through per-load feedback.

In contrast, our setting reverses this dynamic: rewards are not gradually scaled by agent load, but are only triggered when a coalition meets or exceeds an unknown activation threshold. There is no partial feedback—an arm either activates successfully or yields no reward at all. This introduces a fundamentally different coordination and exploration challenge, where agents must jointly infer both the reward distribution and the minimum team size required for success.

Unlike Wang et al.~\shortcite{wang2022multi} or prior work that assumes a known task structure, our agents must learn which arms are promising \textit{and} how many collaborators are needed to activate them. This adds a unique ambiguity to feedback: a zero-reward observation may result from either \textit{insufficient group size} or \textit{stochastic failure}. This makes credit assignment and structural learning significantly more difficult than in classical bandits or resource-avoidance settings. Without coordinated exploration, high-reward arms may remain undiscovered. Our work fills this gap by proposing a setting in which \textit{ cooperation is required just to learn}, not just to optimize.

\nocite{shao2024balanced}


\section{Problem Formulation}

We study the problem of \textbf{cooperative multi-agent multi-armed bandits} (MA-MABs), where a team of $M$ agents interacts with a set of $K$ stationary stochastic arms over discrete time steps. Each arm offers an uncertain reward outcome determined by a fixed but unknown success probability. Unlike classical bandit settings, arms in this framework may \textbf{require simultaneous selection by multiple agents} to activate and yield a reward, which we refer to as \textit{threshold-activated rewards}. This setup is visually illustrated in Figure~\ref{fig:uav_coordination}, where UAV agents must learn to form effective coalitions to activate tasks with different coordination requirements.

This cooperative MA-MAB formulation captures fundamental challenges arising in real-world multi-agent systems, such as collaborative robotics, decentralized search-and-rescue, and distributed sensing, where agents must coordinate under uncertainty, often without centralized supervision. A key challenge in these settings is \textit{efficient team allocation}: for example, if two agents are required to carry a heavy object, sending a third agent achieves the task but wastes the extra agent's effort. Overcommitted agents represent a lost opportunity cost, as they could have been assigned to assist with other valuable tasks.

Thus, the objective is to design decentralized learning strategies that enable agents to jointly explore and exploit the environment, maximizing cumulative team reward over time while minimizing \textbf{team regret}---the gap between the team's achieved reward and the reward attainable with perfect knowledge.

\paragraph{Formal Model.}
There are $M$ agents and $K$ arms, indexed by $i=1, \dots, K$. Time proceeds in discrete rounds $t=1,2,\dots, T$. At each round, each agent selects an arm to pull. The agents' selections collectively define a \textit{joint action} at round $t$, which determines the coalition sizes $N_{i,t}$ for each arm $i$.

Each arm $i$ is associated with two unknown properties:
\begin{itemize}
    \item A threshold $h_i \in \{1, \dots, M\}$, representing the minimum number of agents that must pull arm $i$ simultaneously to activate a reward.
    \item A stochastic reward process, with reward magnitude $r_i$ and success probability $p_i$.
\end{itemize}
Specifically, the reward generation follows a Bernoulli process: when a coalition of agents satisfy the threshold $h_i$, a Bernoulli trial with success probability $p_i$ determines whether the reward $r_i$ is realized and shared among the participating agents, reflecting cooperative tasks where joint success benefits all contributors equally.

The environment behaves as follows:
\begin{itemize}
    \item If $N_{i,t} \geq h_i$, then with probability $p_i$, the reward $r_i$ is realized and split equally among the participating agents.
    \item If $N_{i,t} < h_i$, no reward is given for arm $i$.
\end{itemize}

Agents observe only their individual received rewards after each round; they do not observe the true thresholds $h_i$ or success probabilities $p_i$ directly. Agents may communicate locally to share observations at each round. We assume that communication between agents is cost-free and instantaneous.

\paragraph{Objective and Regret.}
The goal is to maximize the cumulative team reward over time through decentralized learning and coordinated action. Equivalently, the performance is measured via the cumulative \textit{team regret} over $T$ rounds, defined as:

\[
\text{Regret}(T) = T \cdot \mu^* - \mathbb{E}[R_{\text{team}}(T)],
\]
where $\mu^*$ denotes the expected team reward per round under the optimal joint action.

The cumulative team reward up to round $T$ is defined as:
\[
R_{\text{team}}(T) = \sum_{t=1}^{T} \sum_{i=1}^{M} r_{i,t},
\]
where $r_{i,t}$ denotes the reward received by agent $i$ at round $t$, given by:
\[
r_{i,t} =
\begin{cases}
\frac{r_j}{N_{j,t}} & \text{if } a_{i,t} = j,\, N_{j,t} \geq h_j, \, \text{success}, \\
0 & \text{otherwise},
\end{cases}
\]
where $a_{i,t}$ is the action chosen by agent $i$ at round $t$, $N_{j,t}$ denotes the number of agents selecting arm $j$, $h_j$ is the unknown activation threshold, and $r_j$ is the reward magnitude. Success is determined by a Bernoulli trial with probability $p_j$.

\paragraph{Challenges.}
This setting introduces several key challenges:
\begin{itemize}
    \item \textbf{Threshold Uncertainty:} Agents must learn the unknown activation thresholds $h_i$ for each arm.
    \item \textbf{Reward Uncertainty:} Agents must simultaneously estimate expected rewards for each arm under thresholded activation.
    \item \textbf{Decentralized Coordination:} Agents must coordinate to form coalitions that meet thresholds without centralized control.
    \item \textbf{Partial Observability:} Agents only observe their own rewards and must share information effectively.
\end{itemize}

This motivates the development of Threshold-Coop-UCB, a decentralized learning algorithm that jointly estimates thresholds and rewards while forming dynamic coalitions.

\vspace{0.5em}
\noindent

\nocite{bubeck2012regret}

\section{Threshold-Coop-UCB Algorithm}

We propose \textbf{Threshold-Coop-UCB} (T-Coop-UCB), a decentralized learning algorithm that enables a team of agents to jointly estimate unknown activation thresholds and reward distributions while coordinating their actions to maximize cumulative team reward. T-Coop-UCB leverages synchronized global estimates without centralized action assignment; all agents independently select actions using a shared policy based on synchronized estimates of arm statistics and thresholds. Thus, decisions remain entirely decentralized, with no single agent assigning actions or controlling others.

T-Coop-UCB builds on the classical UCB framework introduced by Auer et al.~\shortcite{auer2002finite}, incorporating two mechanisms to address the structural uncertainty in our setting:
\begin{itemize}
    \item \textbf{Threshold Estimation:} Agents maintain estimates $\hat{h}_i(t)$ of the minimum number of agents required to activate each arm. These are updated based on observed coalition sizes and reward feedback, enabling agents to learn the latent coordination requirements.
    
    \item \textbf{Reward Estimation:} Following the standard UCB approach, agents maintain sample-mean reward estimates $\hat{\mu}_i(t)$ for each arm and compute upper confidence bounds to balance exploration and exploitation.
\end{itemize}

\paragraph{Assumptions.}
Threshold-Coop-UCB operates under the following assumptions:  
(1) each arm’s success probability $p_i$ and activation threshold $h_i$ are fixed and stationary over time;  
(2) agents observe only their own individual rewards after each round;  
(3) communication between agents is instantaneous and cost-free, allowing decentralized agents to access consistent global state estimates at each round; while agents independently make their decisions, relaxing this communication assumption presents a meaningful direction for future investigation;
(4) rewards are generated via independent Bernoulli trials conditioned on satisfying the arm’s threshold; and  
(5) coordination remains fully decentralized—no centralized controller assigns actions or mediates decisions.


At each round $t$, agents perform the following steps:
\begin{enumerate}
    \item \textbf{Compute UCB Values.}  
    For each arm $i$, compute:
    \[
    \text{UCB}_i(t) = \hat{\mu}_i(t) + \sqrt{\frac{2 \log t}{n_i(t)}},
    \]
    where $n_i(t)$ is the number of successful threshold-satisfying pulls of arm $i$.
    
    \item \textbf{Coalition Formation.}  
    Agents greedily assign themselves to arms to maximize the sum of UCB values while meeting the estimated thresholds $\hat{h}_i(t)$. Remaining unassigned agents are either allocated to the next-best available arms, if feasible, or remain idle for the round.

    \item \textbf{Action Execution.}  
    Agents pull assigned arms and observe individual rewards, updating estimates only for successfully activated arms.

    \item \textbf{Threshold and Reward Updates.}
    \begin{itemize}
        \item \textbf{Threshold Updates:} Threshold estimates are adjusted conservatively, requiring $m$ consecutive failures at a given coalition size before increasing $\hat{h}_i(t)$. If an arm succeeds with fewer agents than estimated, $\hat{h}_i(t)$ is decreased.
        \item \textbf{Reward Updates:} For arms that succeed, agents update the sample-mean reward estimate $\hat{\mu}_i(t)$ incrementally.
    \end{itemize}
\end{enumerate}

The full T-Coop-UCB procedure is presented in Algorithm~\ref{alg:tcoopucb}.

\begin{algorithm}[tb]
\caption{Threshold-Coop-UCB (T-Coop-UCB)}
\label{alg:tcoopucb}
\begin{algorithmic}[1] 
\STATE \textbf{Input}: Number of agents $M$, number of arms $K$, time horizon $T$
\STATE \textbf{Parameters}: Failure threshold $m$ (for threshold updates)
\STATE \textbf{Output}: $\text{Cumulative team reward } R_{\text{team}}(T)$
\STATE Initialize $\hat{\mu}_i \leftarrow 0$, $\hat{h}_i \leftarrow M$, $n_i \leftarrow 0$ for all arms $i$
\FOR{each round $t=1,2,\dots,T$}
    \STATE Agents communicate and synchronize local estimates
    \FOR{each arm $i$}
        \STATE Compute UCB: $\text{UCB}_i(t) = \hat{\mu}_i(t) + \sqrt{2 \log t / n_i(t)}$
    \ENDFOR
    \STATE Sort arms by descending UCB scores
    \STATE Assign agents greedily to arms based on $\hat{h}_i(t)$
    \STATE Agents pull assigned arms and observe rewards
    \FOR{each arm $i$ pulled}
        \IF{Success with fewer agents than $\hat{h}_i(t)$}
            \STATE Decrease $\hat{h}_i(t)$
        \ELSIF{Failure after $m$ attempts at coalition size}
            \STATE Increase $\hat{h}_i(t)$
        \ENDIF
        \IF{Successful activation}
            \STATE Update $\hat{\mu}_i(t)$ using incremental sample average
            \STATE Increment $n_i(t)$
        \ENDIF
    \ENDFOR
\ENDFOR
\STATE \textbf{return} Cumulative team reward $R_{\text{team}}(T)$
\end{algorithmic}
\end{algorithm}

\section{Experimental Setup}

We simulate environments with $M$ agents and $K$ stationary stochastic arms, each associated with a success probability $p_i$, reward magnitude $r_i$, and activation threshold $h_i$. Arm thresholds and reward distributions are initially unknown to the agents, requiring distributed exploration and decentralized learning.

We focus on scenarios where $M > K$, meaning the number of agents exceeds the number of arms. This assumption shifts the problem from classical constrained resource allocation toward a \textit{cooperative multi-agent learning problem}, where agents must coordinate to form coalitions, jointly learn which arms are profitable, and allocate themselves across arms to maximize cumulative team reward.

For added complexity, some arms are designed as \textit{decoys}: they may require any number of agents to activate but have a success probability $p_j = 0$ or a reward $r_j = 0$, resulting in zero payoff even if the activation threshold is satisfied. This introduces ambiguity in the feedback signal—when agents observe a reward of zero, it is unclear whether the failure was due to stochastic randomness or insufficient coalition size. This ambiguity makes credit assignment and structural learning significantly more difficult than in classical bandits or collision-aware MABs. Agents must therefore not only coordinate effectively, but also identify and avoid such traps to optimize long-term team performance.

\paragraph{Environment.}
Each environment consists of $M=3$ agents and $K=5$ arms. Agents are assumed to be homogeneous in capability and behavior, and no roles or specializations are assigned. All agents apply the same decentralized policy and have equal ability to pull any arm. Thresholds (required number of agents) $h_i$ vary across arms, with some arms requiring coalitions of multiple agents to activate ($h_i > 1$) and others allowing individual activation ($h_i = 1$). Rewards are generated via independent Bernoulli trials conditioned on successful threshold activation. When an arm succeeds, the reward $r_i$ is divided equally among all participating agents. In this environment, the optimal strategy is for all three agents to coordinate on Arm 2 at each round, as it offers the highest expected team reward given its activation threshold of three agents $h_2=3$ and success probability $p_2=0.6$.
 The base environment configuration is summarized in Table~\ref{tab:env_config}.

\begin{table}[tb]
\centering
\small
\caption{Arm configuration for the base environment.}
\label{tab:env_config}
\begin{tabular}{lrrr}
\toprule
Arm $i$ & Success Prob. $p_i$ & Reward $r_i$ & Req. Agents $h_i$ \\
\midrule
0 & 0.5 & 5.0 & 1 \\
1 & 0.7 & 6.0 & 1 \\
2 & 0.6 & 20.0 & 3 \\
3 & 0.4 & 12.0 & 2 \\
4 (Decoy) & 0.0 & 0.0 & 2 \\
\bottomrule
\end{tabular}
\end{table}

\paragraph{Experimental Parameters.}
All experiments are averaged over $30$ independent runs. Agents communicate cost-free at every round to synchronize local observations. The time horizon is set to $T=10,000$ rounds, ensuring long-term distributed learning.

\paragraph{Baselines Compared.}
We compare T-Coop-UCB against:
\begin{itemize}
    \item \textbf{Random Policy}: Agents choose arms uniformly at random, without cooperation or learning.
    \item \textbf{Independent UCB1}: Each agent independently applies UCB1 without communication, cooperation, or coordinated coalition formation.
    \item \textbf{Cooperative UCB1}: Agents share observations and coordinate actions under known thresholds but without adaptive threshold learning. 
    \item \textbf{Oracle}: A centralized policy with full knowledge of thresholds and reward distributions. At each round, it selects the joint action that maximizes expected team reward, serving as a performance upper bound.

\end{itemize}

\paragraph{Evaluation Metrics.}
We evaluate performance using several complementary metrics:

\begin{itemize}
    \item \textbf{Cumulative Team Reward:} Total team reward accumulated over $T$ rounds.
    \item \textbf{Cumulative Team Regret:} Difference between the cumulative reward of an Oracle policy with perfect knowledge and the team's achieved reward.
    \item \textbf{Valid Allocations:} Number of rounds where the coalition assigned to an arm met or exceeded its required threshold $h_i$, regardless of stochastic success.
\end{itemize}

\section{Results and Analysis}

We evaluate the performance of T-Coop-UCB compared to Random Policy, Independent UCB1, Cooperative UCB1, and a centralized policy with full knowledge (Oracle) across several metrics: cumulative team reward, team regret, and valid allocations.

\begin{figure*}[h]
    \centering
    \includegraphics[width=\textwidth]{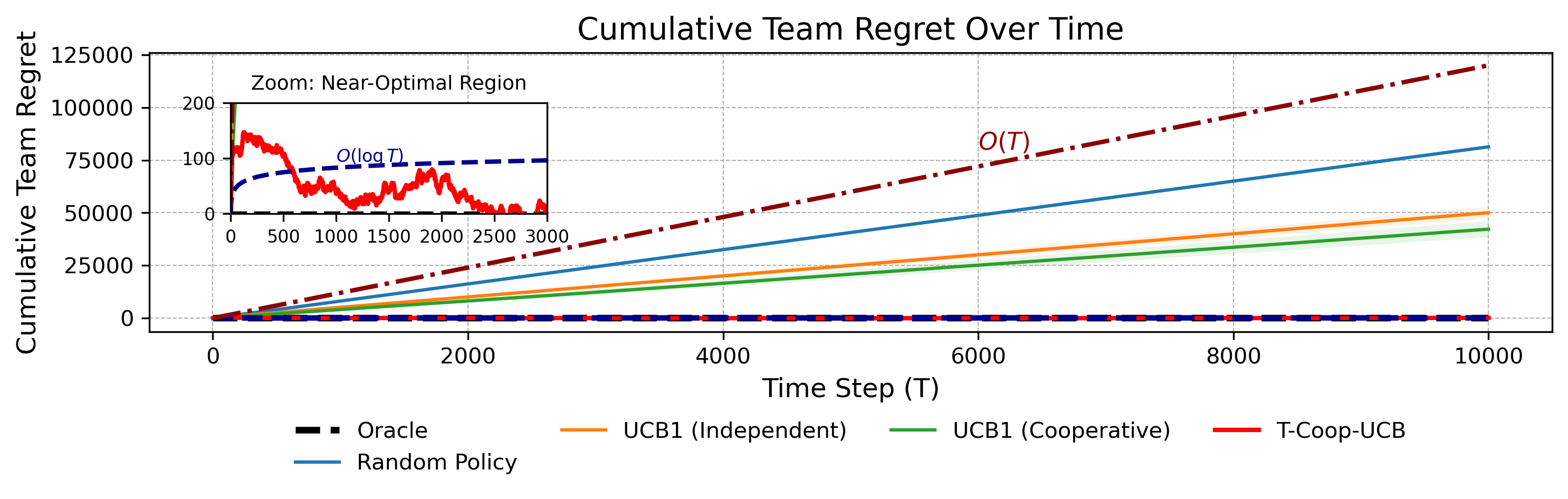}
    \caption{\textbf{Cumulative team regret over time.} T-Coop-UCB rapidly approaches near-Oracle performance and consistently achieves lower regret than all baselines. While all learning-based methods exhibit sublinear regret, T-Coop-UCB empirically tracks below the $O(\log T)$ reference, indicating highly efficient learning. Shaded areas show 95\% confidence intervals over 30 runs. Theoretical baselines ($O(T)$, $O(\log T)$) are scaled by the optimal expected reward ($\mu^* = 12$).}

    \label{fig:cumulative_regret_linear}
\end{figure*}

\paragraph{Cumulative Team Regret.}
We focus on cumulative team regret as the primary metric, which reflects how efficiently each policy learns to coordinate under structural uncertainty. As shown in Figure~\ref{fig:cumulative_regret_linear}, all learning-based baselines achieve sublinear regret, outperforming the naive $O(T)$ benchmark. Among them, T-Coop-UCB exhibits the lowest regret and converges rapidly toward the Oracle’s performance. Although Cooperative UCB1 benefits from knowing the true thresholds, it lacks the flexibility to adapt under ambiguous feedback. When an arm fails to yield a reward, Cooperative UCB1 cannot distinguish whether the cause is stochastic noise or insufficient coordination, leading it to lower its confidence estimate and avoid revisiting potentially valuable arms.

T-Coop-UCB, by contrast, learns from ambiguous signals. It gradually infers both activation thresholds and reward distributions through repeated interaction, refining agent allocation strategies over time. Despite the added complexity of decentralized learning, it maintains low regret growth. This mirrors the $O(\log T)$ regret bounds established in cooperative multi-agent bandits with known structure~\cite{landgren2016distributed}, suggesting that T-Coop-UCB achieves similar efficiency even when coordination requirements must be learned. Interestingly, T-Coop-UCB’s regret curve often tracks below the $O(\log T)$ benchmark, although it exhibits mild oscillations due to the cost of joint exploration and temporary miscoordination. We view this as a reflection of the underlying structural learning challenge, and a promising direction for future work, such as smoothing techniques or adaptive mechanisms to improve convergence stability. While we do not claim theoretical sub-logarithmic regret, this empirical behavior highlights the algorithm’s ability to uncover task structure early, enabling faster convergence and fewer costly misallocations.

\begin{figure}[tb]
    \centering
    \includegraphics[width=\linewidth]{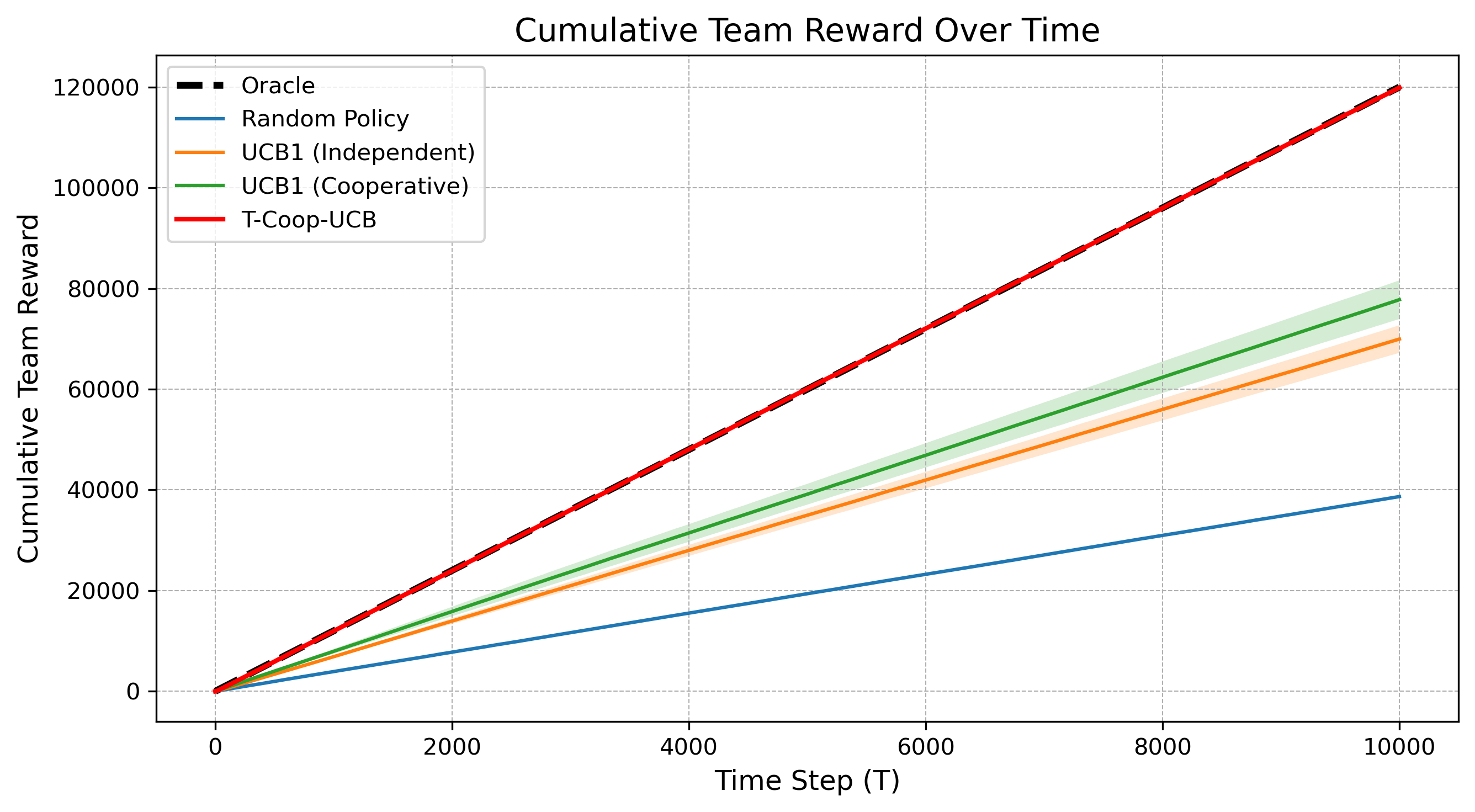}
    \caption{\textbf{Cumulative team reward over time.} T-Coop-UCB rapidly converges to near-optimal team performance, closely tracking the Oracle policy. Shaded regions represent 95\% confidence intervals over 30 independent runs. }

    \label{fig:cumulative_reward}
\end{figure}

\paragraph{Cumulative Team Reward.}
Figure~\ref{fig:cumulative_reward} shows that T-Coop-UCB achieves substantially higher cumulative team reward than both Independent and Cooperative UCB1, closely tracking the Oracle policy throughout learning. This indicates that T-Coop-UCB effectively learns both the reward distributions and the hidden activation thresholds necessary to reliably activate high-reward arms (e.g., Arm 2).

Without the ability to estimate activation thresholds, the baseline strategies underperform. Cooperative UCB1 outperforms Independent UCB1 due to agents' ability to share reward observations, reinforcing prior findings that information sharing improves decentralized decision quality~\cite{chakraborty2017coordinated,landgren2021distributed}. However, despite having access to the true thresholds, Cooperative UCB1 still underperforms relative to T-Coop-UCB due to its inability to adaptively explore and disambiguate between stochastic failures and threshold violations. When an arm fails once or twice, Cooperative UCB1 lowers its UCB estimate and prematurely abandons potentially valuable arms, lacking the structural feedback interpretation and robustness that T-Coop-UCB incorporates. 

Finally, Independent UCB1, with its greedy arm selection strategy~\cite{auer2002finite}, outperforms Random Policy, but lacks communication and joint coordination, leading to frequent failure to meet activation thresholds of high-reward arms. As expected, Random Policy accumulates the least reward, highlighting the necessity of both structured exploration and coordinated action in threshold-activated environments.

\begin{figure}[tb]
    \centering
    \includegraphics[width=0.8\linewidth]{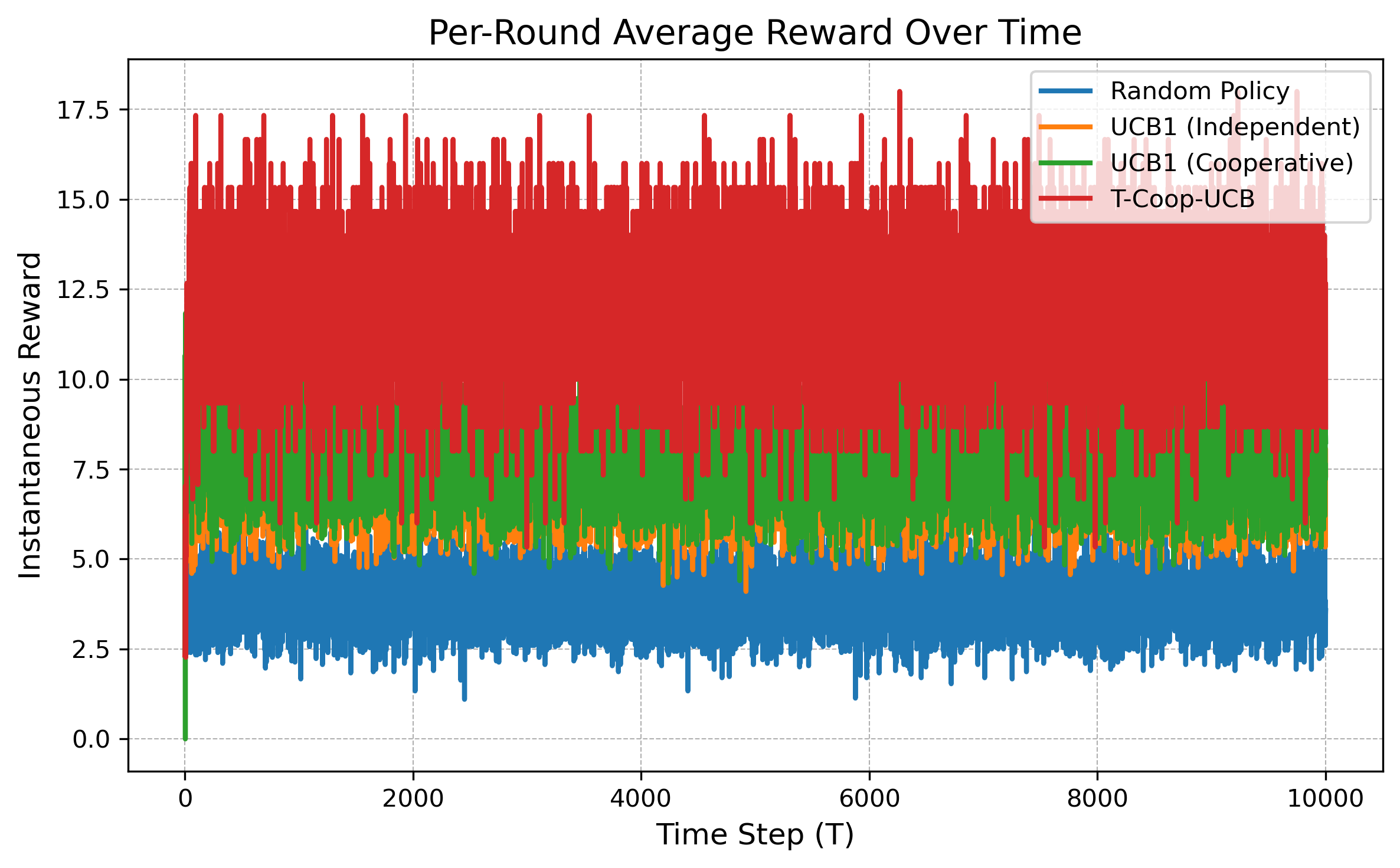}
    \caption{\textbf{Per-round average reward over time.} T-Coop-UCB rapidly converges to high average reward by successfully exploiting high-value arms, while baselines without threshold learning or coordination lag behind.}

    \label{fig:instantaneous_reward}
\end{figure}

\paragraph{Per-Round Average Reward.}
Figure~\ref{fig:instantaneous_reward} shows that T-Coop-UCB achieves a consistently higher average reward per round than all baselines. This reflects the algorithm's ability to learn both reward distributions and coalition sizes more effectively than strategies without threshold adaptation or coordination.


\begin{figure}[tb]
    \centering
    \includegraphics[width=0.8\linewidth]{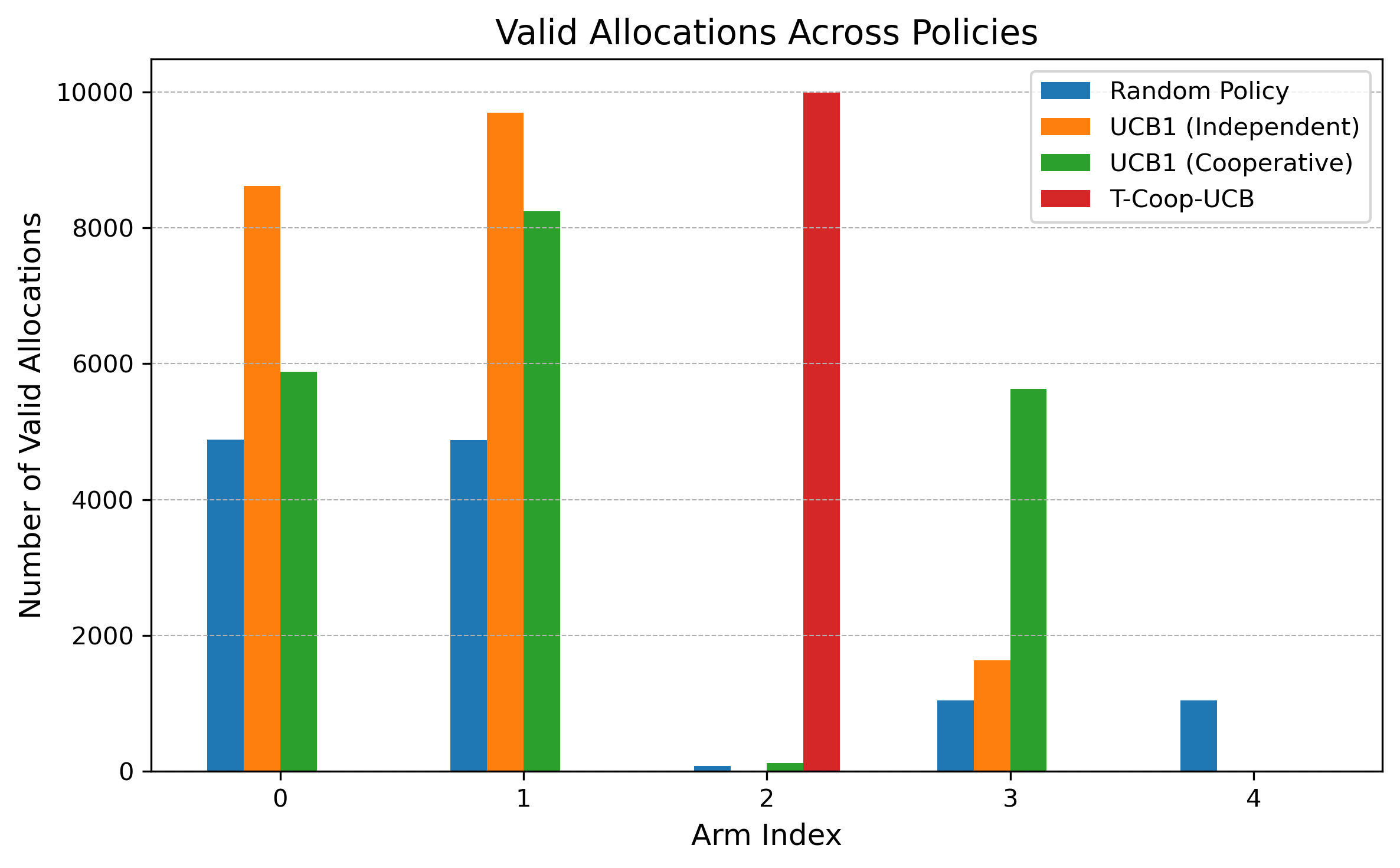}
    \caption{\textbf{Valid allocations per arm across policies.} An allocation is valid when the number of agents meets or exceeds an arm’s activation threshold. T-Coop-UCB consistently forms effective coalitions, outperforming baseline methods.}

    \label{fig:valid_allocations}
\end{figure}

\paragraph{Valid Arm Allocation and Reward Success.}
Figure~\ref{fig:valid_allocations} shows that T-Coop-UCB consistently learns to assign the right number of agents to high-value arms, forming successful coalitions even under uncertainty. In contrast, baseline policies often struggle to coordinate, especially on arms that require teamwork, underscoring how difficult decentralized learning can be in structured environments. T-Coop-UCB not only allocates agents more effectively, but also activates high-reward arms more often, particularly Arm 2, whereas baselines tend to avoid these arms due to limited coordination or an inability to recognize the threshold requirements.

Over time, T-Coop-UCB identifies Arm 2 to be the most valuable and that it takes all three agents for activation. Once learned, it reliably sends the full team to that arm. Cooperative UCB1 explores Arm 2 early on but lacks the ability to interpret failure, namely its unable to tell whether the reward did not trigger because of bad luck or too few agents. As a result, it retreats to easier arms with lower thresholds like Arms 0, 1, and 3. Meanwhile, Independent UCB1 and Random Policy, which lack coordination, perform best on arms that only require one agent, reinforcing the importance of communication and teamwork in this setting.

Arm 4, designed as a decoy with no reward, draws the most attention from Random Policy simply by chance. Because it explores uniformly, it occasionally sends enough agents to meet the threshold. In contrast, learning-based policies like T-Coop-UCB and Cooperative UCB1 quickly realize that the arm has no payoff and learn to avoid it, demonstrating their ability to filter out dead ends through shared feedback and structured learning.

\section{Discussion and Future Work}

\paragraph{Discussion.}
Our results show that learning activation thresholds is critical for decentralized cooperation: algorithms that do not adapt to unknown coordination requirements, such as Independent UCB1 or Random Policy, perform poorly, and even methods with access to true thresholds, like Cooperative UCB1, underperform without joint reward learning.  We demonstrate that T-Coop-UCB effectively learns both activation thresholds and reward distributions in decentralized cooperative multi-agent bandit settings. Our algorithm consistently outperforms baselines across all metrics, achieving near-Oracle cumulative reward and significantly lower cumulative  team regret. These results highlight the significance of distributed threshold learning in enabling efficient coalition formation and maximizing team performance.

While T-Coop-UCB achieves strong performance, several assumptions simplify the current setting. In particular, we assume perfect cost-free communication among agents and stationary environments with fixed thresholds and reward probabilities. Additionally, the environments considered involve relatively small numbers of agents and arms, with full observability simulated via communication in the cooperative settings.

\paragraph{Future Work.}
While our algorithm empirically demonstrates strong performance, formal theoretical guarantees regarding regret bounds and threshold estimation remain open problems. Providing such theoretical analysis under structured assumptions (e.g., stationary environments, known reward distributions) is an important avenue for future research. Additionally, our current empirical evaluations focus on relatively small-scale scenarios. Exploring scalability, along with robustness under communication noise, dynamic thresholds, or partial observability, is left for future study.

Beyond these foundational analyses, this work opens several promising avenues for further research. One direction involves extending T-Coop-UCB to settings with communication constraints or intermittent synchronization, where agents must learn effectively under partial observability and delayed information exchange. Another is adapting the algorithm to non-stationary environments where activation thresholds and reward distributions evolve over time. Addressing scalability to support larger agent populations and arm sets, while preserving decentralized coordination efficiency, remains an important challenge. Finally, future work could explore more complex coordination scenarios, including heterogeneous task requirements, dynamic task arrival, and adaptive coalition formation.

\section{Conclusion}


We explore how teams of autonomous agents can learn to make cooperative decisions under uncertainty, particularly when tasks require joint, coordinated actions. We model this as a decentralized multi-agent bandit problem and introduce Threshold-Coop-UCB (T-Coop-UCB), an algorithm that enables agents to learn both how many teammates are needed to activate a task and whether the task is worth pursuing. By jointly estimating activation thresholds and rewards, agents form efficient coalitions that maximize long-term team performance.

This addresses a key gap in prior work, which often assumes known coordination structures or independently triggered rewards. We present a setting with strong real-world relevance, where agents must discover both whom to coordinate with and which tasks are valuable—all through experience. We also introduce \textit{decoy arms}, which require coordination to activate but yield no reward, complicating joint exploration by introducing structural traps. This challenge reflects scenarios like UAV-based disaster response and distributed robotics, where success depends on adapting to unknown and misleading task requirements in real time.

Our results show that T-Coop-UCB consistently outperforms baseline methods, achieving near-Oracle cumulative reward and rapid regret convergence. These findings underscore the importance of decentralized systems learning not just what to do, but how to work together. Future work includes extending the algorithm to settings with communication constraints, non-stationary dynamics, larger teams, and formal regret guarantees.

\bibliographystyle{named}
\bibliography{ijcai25}

\end{document}